\documentstyle[12pt]{article}
\makeatletter
\def\1ad{\mbox{\normalsize $^1$}}
\def\makefront{\vspace*{1cm}\begin{center}
\def\newtitleline{\\ \vskip 5pt}
{\Large\bf\titleline}\\
\vskip 1truecm
{\large\bf\authors}\\
\vskip 5truemm
\addresses
\end{center}
\vskip 1truecm
{\bf Abstract:}
\abstracttext
\vskip 1truecm}
\newdimen\normalarrayskip              % skip between lines
\newdimen\minarrayskip                 % minimal skip between lines
\normalarrayskip\baselineskip
\minarrayskip\jot
\newif\ifold             \oldtrue            \def\new{\oldfalse}
\def\arraymode{\ifold\relax\else\displaystyle\fi} % mode of array entries
\def\eqnumphantom{\phantom{(\theequation)}}     % right phantom in eqnarray
\def\@arrayskip{\ifold\baselineskip\z@\lineskip\z@
     \else
     \baselineskip\minarrayskip\lineskip2\minarrayskip\fi}
\def\@arrayclassz{\ifcase \@lastchclass \@acolampacol \or
\@ampacol \or \or \or \@addamp \or
   \@acolampacol \or \@firstampfalse \@acol \fi
\edef\@preamble{\@preamble
  \ifcase \@chnum
     \hfil$\relax\arraymode\@sharp$\hfil
     \or $\relax\arraymode\@sharp$\hfil
     \or \hfil$\relax\arraymode\@sharp$\fi}}
\def\@array[#1]#2{\setbox\@arstrutbox=\hbox{\vrule
     height\arraystretch \ht\strutbox
     depth\arraystretch \dp\strutbox
     width\z@}\@mkpream{#2}\edef\@preamble{\halign
\noexpand\@halignto
\bgroup \tabskip\z@ \@arstrut \@preamble \tabskip\z@ \cr}%
\let\@startpbox\@@startpbox \let\@endpbox\@@endpbox
  \if #1t\vtop \else \if#1b\vbox \else \vcenter \fi\fi
  \bgroup \let\par\relax
  \let\@sharp##\let\protect\relax
  \@arrayskip\@preamble}
\def\eqnarray{\stepcounter{equation}%
              \let\@currentlabel=\theequation
              \global\@eqnswtrue
              \global\@eqcnt\z@
              \tabskip\@centering
              \let\\=\@eqncr
              $$%
 \halign to \displaywidth\bgroup
    \eqnumphantom\@eqnsel\hskip\@centering
    $\displaystyle \tabskip\z@ {##}$%
    &\global\@eqcnt\@ne \hskip 2\arraycolsep
         $\displaystyle\arraymode{##}$\hfil
    &\global\@eqcnt\tw@ \hskip 2\arraycolsep
         $\displaystyle\tabskip\z@{##}$\hfil
         \tabskip\@centering
    &{##}\tabskip\z@\cr}
\makeatother

\begingroup\ifx\undefined\newsymbol \else\def\input#1 {\endgroup}\fi
\input amssym.def \relax
\input amssym
\newfont{\hr}{msbm10}
\newfont{\ams}{msam10}

\def\*{\star}
\def\({\left(}		
\def\){\right)}		
\def\[{\left[}		
\def\]{\right]}

\def\frac#1#2{{#1 \over #2}}

\def\dsl{\raise.15ex\hbox{/}\kern-.57em\partial}
\def\Dsl{\,\raise.15ex\hbox{/}\mkern-.13.5mu D}
\def\be{\beta}

\font\numbers=cmss12
\font\upright=cmu10 scaled\magstep1
\def\stroke{\vrule height8pt width0.4pt depth-0.1pt}
\def\topfleck{\vrule height8pt width0.5pt depth-5.9pt}
\def\botfleck{\vrule height2pt width0.5pt depth0.1pt}
\def\Zmath{\vcenter{\hbox{\numbers\rlap{\rlap{Z}\kern 0.8pt\topfleck}\kern
2.2pt
                   \rlap Z\kern 6pt\botfleck\kern 1pt}}}
\def\Qmath{\vcenter{\hbox{\upright\rlap{\rlap{Q}\kern
                   3.8pt\stroke}\phantom{Q}}}}
\def\Nmath{\vcenter{\hbox{\upright\rlap{I}\kern 1.7pt N}}}
\def\Cmath{\vcenter{\hbox{\upright\rlap{\rlap{C}\kern
                   3.8pt\stroke}\phantom{C}}}}
\def\Rmath{\vcenter{\hbox{\upright\rlap{I}\kern 1.7pt R}}}
\def\Z{\ifmmode\Zmath\else$\Zmath$\fi}
\def\Q{\ifmmode\Qmath\else$\Qmath$\fi}
\def\N{\ifmmode\Nmath\else$\Nmath$\fi}
\def\C{\ifmmode\Cmath\else$\Cmath$\fi}
\def\R{\ifmmode\Rmath\else$\Rmath$\fi}

\def\Tr{{\rm Tr}}

\def\2{{1\over 2}}

\def\bea{\begin{eqnarray}}
\def\eea{\end{eqnarray}}
\def\nn{\nonumber}
\def\beq{\begin{equation}}
\def\eeq{\end{equation}}
\def\ba{\beq\new\begin{array}{c}}
\def\ea{\end{array}\eeq}
\def\be{\ba}
\def\ee{\ea}
\def\N2{${\cal N}=2$}

\begin {document}
\begin{flushright}
FIAN/TD-02/98\\
ITEP/TH-04/98
\end{flushright}
\vspace{-1.2cm}
\def\titleline{
%%%%%%%%%%%%%%%%%%%%%%%%%%%%%%%%%%%%%%%%%%%%%%
%                                            %
% Insert now the text of your title.         %
% Make a linebreak in the title with         %
%                                            %
%            \newtitleline                   %
%                                            %
%%%%%%%%%%%%%%%%%%%%%%%%%%%%%%%%%%%%%%%%%%%%%%
On N=2 SUSY gauge theories
\newtitleline
and integrable systems
%%%%%%%%%%%%%%%%%%%%%%%%%%%%%%%%%%%%%%%%%%%%%%
}
\def\authors{
%%%%%%%%%%%%%%%%%%%%%%%%%%%%%%%%%%%%%%%%%%%%%%
%                                            %
%  Insert now the name (names) of the author %
%  (authors).                                %
%  In the case of several authors with       %
%  different addresses use e.g.              %
%                                            %
%             \1ad          etc.             %
%                                            %
%  to indicate that a given author has the   %
%  address number 1 , 2 , etc.               %
%                                            %
%%%%%%%%%%%%%%%%%%%%%%%%%%%%%%%%%%%%%%%%%%%%%%
A.Mironov \1ad
%%%%%%%%%%%%%%%%%%%%%%%%%%%%%%%%%%%%%%%%%%%%%
}
\def\addresses{
%%%%%%%%%%%%%%%%%%%%%%%%%%%%%%%%%%%%%%%%%%%%%%
%                                            %
% List now the address. In the case of       %
% several addresses list them by numbers     %
% using e.g.                                 %
%                                            %
%             \1ad                           %
%                                            %
% to numerate address 1 , 2 , etc.           %
%                                            %
%%%%%%%%%%%%%%%%%%%%%%%%%%%%%%%%%%%%%%%%%%%%%%
\1ad
Theory Dept., Lebedev Physical Institute, Moscow, Russia\\
and ITEP, Moscow, Russia
%%%%%%%%%%%%%%%%%%%%%%%%%%%%%%%%%%%%%%%%%%%%%%%
}
\def\abstracttext{
%%%%%%%%%%%%%%%%%%%%%%%%%%%%%%%%%%%%%%%%%%%%%%%
%                                             %
% Insert now the text of your abstract.       %
%                                             %
%%%%%%%%%%%%%%%%%%%%%%%%%%%%%%%%%%%%%%%%%%%%%%%
This note gives a brief review of the integrable
structures presented in the Seiberg-Witten approach
to the N=2 SUSY gauge theories with emphasize on
the case of the gauge theories with
matter hypermultiplets included (described by spin chains).
The web of different N=2 SUSY theories is discussed.
%%%%%%%%%%%%%%%%%%%%%%%%%%%%%%%%%%%%%%%%%%%%%%%%%%%%%%%%%%%%%%%
}
\makefront
%%%%%%%%%%%%%%%%%%%%%%%%%%%%%%%%%%%%%%%%%%%%%%%%
%                                              %
%  Insert now the remaining parts of           %
%  your article.                               %
%                                              %
%%%%%%%%%%%%%%%%%%%%%%%%%%%%%%%%%%%%%%%%%%%%%%%%
\paragraph{General remarks.}
Since the paper by N.Seiberg and E.Witten \cite{SW}, there have been a lot of
attempts to get better understanding of the structures arising in the
low-energy sector of N=2 SUSY gauge theories. In a sense, the paper \cite{SW}
pointed out the importance of objects completely different from those
typically dealt with in quantum field theory. In particular, one of the
main quantities in the Seiberg-Witten (SW) approach is the prepotential giving
the low-energy effective action of the theory.

One of the constituent parts of arena where the low-energy effective
theory lives is, in accordance with \cite{SW}, a Riemann surface, while a
subspace of the moduli space of Riemann surfaces gives the moduli space of
vacua of the physical theory.
The whole world-sheet of the low-energy theory can be described in terms of
5-brane in M-theory \cite{W,Gor,MMM,GGM1}.

To these counterparts of the field theory objects, one should also add the
analog of the symmetry principle which arises within the SW framework.
Namely, it turns out that the symmetry properties of theory in the low-energy
limit are encoded in the integrable system that underlines the low-energy
dynamics.

The very fact of existence of an integrable system behind SW solution has
been first realized in the paper \cite{GKMMM} that dealt with the pure gauge
theory. Since then, many more examples of the N=2 SUSY theories has been
investigated, and corresponding integrable systems have been revealed
\cite{GMMM1}-\cite{GMMM2}.

The goal of the present short review is to sketch the general scheme of
connection between the SW solutions and integrable systems. We also
describe the correspondence (SUSY theory
$\longleftrightarrow$ integrable system) in concrete examples discussing
what deformations of integrable systems correspond to deformations of
physical theories.

\paragraph{SW anzatz, prepotential and integrable system.}

For the \N2 SUSY gauge theory
the SW anzatz can be {\em formulated} in the following way.
One starts with two {\it bare} spectral curves. One of them,
with a holomorphic 1-form $d\omega $, is elliptic curve (torus)
\be\label{torus}
E_1(\tau):\ \ \ \ y^2 = \prod_{a=1}^3 (x - e_a(\tau)), \ \ \
\sum_{a=1}^3 e_a(\tau) = 0,\ \ \ \ \ d\omega = \frac{dx}{y}\equiv d\xi,
\ee
when the YM theory contains the adjoint matter hypermultiplet,
or its degeneration $\tau
\rightarrow i\infty$ -- the double-punctured sphere (``annulus''):
\be
x\rightarrow w\pm\frac{1}{w},\ \
y \rightarrow w\mp\frac{1}{w},\ \ \ \ \ d\omega = \frac{dw}{w}
\ee
otherwise. In particular, this latter possibility is the case
for the theory with the fundamental matter hypermultiplets.

The second bare spectral curve is also elliptic curve $E_2(\tau')$
or its degenerations depending on the dimension of the
space-time.

In the integrable framework, the two bare spectral curves are related
by the full spectral curve that is just the Riemann surface emerging
within the Seiberg-Witten construction. There are two different types of
integrable system with the corresponding associated full spectral curve.

Integrable systems of the first type which could be naturally called Hitchin
type systems are described as follows. First, one introduces
the Lax operator ${\cal L}(x,y)$ that is defined as a $N\times N$
matrix-valued function
(1-differential) on the first {\it bare} spectral curves. Then, the
{\it full} spectral curve ${\cal C}$ is given by the Lax-eigenvalue equation:
$\det({\cal L}(x,y) - \lambda ) = 0$, where $\lambda$ is given on the second
bare curve. As a result, ${\cal C}$ arises as a
ramified covering over the {\it bare} spectral curves $E_1(\tau)$:
\be\label{speccurve}
{\cal C}:\ \ \ {\cal P}(\lambda; x,y) = 0
\ee
The typical system of this type is the Calogero-Moser system. The Lax
operators in the systems of the first type satisfy linear
Poisson brackets with generally speaking {\it dynamical} elliptic
$r$-matrix \cite{ER}.

On contrary, integrable systems of the second type are characterized by the
quadratic Poisson brackets with the {\it numerical} $r$-matrix (certainly,
quadratic Poisson relations can be easily rewritten as the linear ones with
dynamical $r$-matrix \cite{Skl}). The typical systems of this type are
lattice systems and spin chains \cite{FT}. They are described by $2\times 2$
matrix-valued transfer-matrices $T_N(\xi)$, and the full spectral curve is
given by the equation $\det(T_N(\xi) - w) = 0$, where $w$ is
given on the second bare curve. In fact, only the systems
when at least one of the bare curves is degenerated are investigated in
detail. Therefore, either $w$ is the coordinate on the cylinder, or $\lambda$
is the coordinate on the sphere or cylinder.

The function ${\cal P}$ in (\ref{speccurve})
depends also on parameters (moduli)
$s_I$, parametrizing the moduli space ${\cal M}$. From the
point of view of integrable system, the Hamiltonians
(integrals of motion) are some specific co-ordinates on
the moduli space. From the point of view of gauge theory, the
co-ordinates $s_I$ include $s_i$ -- (the Schur polynomials of) the
adjoint-scalar expectation values $h_k = \frac{1}{k}\langle\Tr \phi^k\rangle$
of the vector ${\cal N}=2$ supermultiplet, as well as $s_\iota = m_\iota$ --
the masses of the hypermultiplets. One associates with the handle of ${\cal
C}$ the gauge moduli and with punctures -- massive hypermultiplets, masses
being residues in the punctures.

The generating 1-form $dS \cong \xi d\omega$ is meromorphic on
${\cal C}$ (the equality modulo total derivatives
is denoted by ``$\cong$''), where $\xi\in E_2$ is associated with one of the
bare curve and $d\omega$ -- with another one $E_1$ (they are related via the
spectral curve equation). In integrable system terms, this form is just the
shorten action "$pdq$" along the non-contractible contours on the Hamiltonian
tori, i.e. related to the symplectic form "$d\xi\wedge d\omega$".

The prepotential is defined in terms of the
cohomological class of $dS$:
\be
a_I = \oint_{A_I} dS, \ \ \ \ \ \
\frac{\partial F}{\partial a_I} = \int_{B_I} dS \nn \\
A_I \circ B_J = \delta_{IJ}.
\label{defprep}
\ee
The cycles $A_I$ include the $A_i$'s wrapping around the handles
of ${\cal C}$ and $A_\iota$'s, going around the singularities
of $dS$.
The conjugate contours $B_I$ include the cycles $B_i$ and the
{\it non-closed} contours $B_\iota$, ending at the singularities
of $dS$ (see \cite{wdvv} for more details).
The integrals $\int_{B_\iota} dS$ are actually divergent, but
the coefficient of divergent part is equal to residue of $dS$
at particular singularity, i.e. to $a_\iota$. Thus, the divergent
contribution to the prepotential is quadratic in $a_\iota$, while
the prepotential is normally defined {\it modulo} quadratic combination
of its arguments (which just fixes the bare coupling constant). In
particular models $\oint_{A,B} dS$ for some conjugate pairs of contours are
identically zero on entire ${\cal M}$: such pairs are not included into our
set of indices $\{I\}$.

Note that the data the period matrix of ${\cal C}$ $T_{ij}(a_i)=
{\partial^2 F\over \partial a_i\partial a_j}$
as a function of the action variables $a_i$
gives the set of coupling constants in the effective theory.

The most important property of the differential $dS\cong\xi{dw\over w}$
is that its derivatives w.r.t. moduli gives holomorphic differentials on
${\cal C}$ (see \cite{wdvv}). The prepotential in the context of integrable
systems was also discussed in \cite{ISWL}.

\paragraph{Spin chains: gauge theories with fundamental matter.}
The crucial difference between integrable systems of the two types is in
interpretation of the spectral curve determinant equation. It is the general
corollary of existence of the linear Poisson bracket (even with dynamical
$r$-matrix) that the spectral determinant equation generates the conserved
quantities \cite{VM}. Therefore, the coefficients of the spectral curve
polynomial are the integrals of motion (and give some coordinates on the
moduli space). However, in integrable system of the second type, there is
some more direct meaning of the spectral determinant equation.

Namely, it can be described as periodicity condition that is imposed onto the
transfer-matrix. In fact, the existence of the transfer-matrix describing the
evolution into the discrete direction (see \cite{FT}) is the main peculiarity
of this kind integrable systems.

The simplest
example of the integrable system of the second type is the periodic Toda
chain. This system is, at the same time, of the first (Hitchin) type. This
surprising fact looks accidental and is due to the possibility of two
different descriptions of the Toda chain \cite{GMMM1}. The Lax operators in
the first description satisfy linear Poisson brackets with the trigonometric
{\it numerical} $r$-matrix, while in the second one -- those with the rational
numerical $r$-matrix.  In this case, $E_1$ degenerates into the cylinder with
coordinate $w$, while $E_2$ -- into the sphere with coordinate
$\lambda$.\footnote{This latter degeneration of the torus $E(x,y)$ is
described as $x\to 0,\ y\to \lambda$, or $x\to\lambda,\ y\to 0$ and
$d\omega\to d\lambda$.}

Let us now switch to the SUSY gauge theories that are underlined by the above
described integrable systems. The periodic Toda chain is associated with the
pure gauge \N2 theory \cite{GKMMM,MW}. Possible deformations of this physical
theory is to add matter hypermultiplets. In fact, one can add either
one matter hypermultiplet in adjoint representation which gives rise to the
UV-finite theory, or several fundamental matter hypermultiplets so that the
theory is still asymptotically free, or UV-finite\footnote{For the $SU(N_c)$
gauge group, there should be at most $2N_c$ fundamental hypermultiplets.}.
This corresponds to the two natural deformations of the periodic Toda chain.
The first deformation (by adding the adjoint matter) is associated with the
elliptic Calogero-Moser system \cite{IntCal} and is the deformation within the
Hitchin-like approach to the Toda system. The other possible way to deform
the Toda chain is to consider more general system admitting the
transfer-matrix description.  This system is the (inhomogeneous) periodic
$XXX$ spin chain and describes exactly the theory with fundamental
matter \cite{GMMM1}.

In fact, there are more general spin chains that can be also associated with
some physical theories. We return to this question in the next paragraph. Now
let us just note that all periodic inhomogeneous chains admit the general
description so that the chain of length $n$ is given by the Lax matrices
$L_i(\xi+\xi_i)$, $\xi_i$ being the chain inhomogeneities, and periodic
boundary conditions. Thus, integrable systems of the second type differ from
each other only by different concrete Lax operators $L_i(\xi)$
\cite{GGM1,GGM2}.

The linear problem in the spin
chain has the following form
\be\label{lproblem}
L_i(\xi)\Psi_i(\xi)=\Psi_{i+1}(\xi)
\ee
where $\Psi_i(\xi)$ is the two-component Baker-Akhiezer function.
The periodic boundary conditions are easily formulated in terms
of the Baker-Akhiezer function and read as
\be\label{pbc}
\Psi_{i+n}(\xi)=-w\Psi_{i}(\xi)
\ee
where $w$ is a free parameter (diagonal matrix).
One can introduce the transfer matrix shifting $i$ to $i+n$
\be\label{Tmat}
T(\xi)\equiv L_n(\xi)\ldots L_1(\xi)
\ee
which provides the spectral curve equation
\be\label{specurv}
\det (T(\xi)+w\cdot {\bf 1})=0
\ee
and generates a complete set of integrals of motion. Note that  in this
approach the parameter $w$ of the bare spectral curve $E_1$ just describes
the periodicity conditions.

Integrability of the spin chain follows from
{\it quadratic} r-matrix relations (see, e.g. \cite{FT})
\be\label{quadr-r}
\left\{L_i(\xi)\stackrel{\otimes}{,}L_j(\xi')\right\} =
\delta_{ij}
\left[ r(\xi-\xi'),\ L_i(\xi)\otimes L_i(\xi')\right]
\ee

The crucial property of this relation is that it
is multiplicative and any product like (\ref{Tmat})
satisfies the same relation
\be\label{Tbr}
\left\{T(\xi)\stackrel{\otimes}{,}T(\xi')\right\} =
\left[ r(\xi-\xi'),\
T(\xi)\otimes T(\xi')\right]
\ee

\paragraph{Zoo of \N2 SUSY theories}
Thus far, we mentioned two possible generalizations of the periodic Toda
chain: to the elliptic Calogero-Moser system, and to the inhomogeneous $XXX$
spin chain. In fact, these integrable systems admit further
deformations. Indeed, the elliptic Calogero-Moser system is the system with
coordinate variables living on the torus, but momentum variables -- on the
sphere.  One can naturally deform this system to involve momenta living on
the cylinder (elliptic Ruijsenaars model \cite{ERu}) or even on another torus
(the second elliptic bare curve, double elliptic system \cite{fgnr}).

At the same time, $XXX$ spin chains described by the rational $r$-matrix can
be deformed to the either $XXZ$ or $XYZ$ spin chains, described
correspondingly by the trigonometric or elliptic $r$-matrices
\cite{FT}. In these theories the second bare curve is the
cylinder or torus respectively, and this is the manifold where momenta of the
system lives. It implies that the momentum variables of integrable system get
restricted values. One might think of this as of a sort of Kaluza-Klein
mechanism. This interpretation, indeed, turns out to be correct so that the
$XXZ$ spin chain describes the 5 dimensional SUSY gauge theories with
fundamental matter\footnote{The pure gauge theory in 5d is described by
degeneration of the $XXZ$ chain, namely, by the relativistic Toda chain
\cite{nikita,wdvv,GGM2}.} and one of the 5 dimensions compactified onto the
circle \cite{GGM2,MM}.  At the same time, the $XYZ$ chain describes the 6
dimensional theory with fundamental matter and with 2 dimensions compactified
onto the bare torus \cite{GMMM2,GGM2,MM}.

Note that 6 dimensions exhaust the room for consistent theories. It perfectly
matches the fact that the $XYZ$ chain seems not to admit further
deformations. Still there is yet another possible deformation of the spin
chain. Namely, one can consider, instead of $sl(2)$, (inhomogeneous) $sl(p)$
spin chains that are described by the $p\times p$ matrix-valued Lax operators
\cite{FT,GGM1,GGM2}. Such systems are associated \cite{GGM1} with the SUSY
theory with the gauge group being the product of simple factors and with
bi-fundamental matter hypermultiplets \cite{W}.

After these identifications made, one can build the whole picture of the
correspondence (integrable systems $\longleftrightarrow$ SUSY gauge
theories). As the starting point, one describes the pure gauge theory
by the simple periodic Toda chain. Then, adding matter hypermultiplets leads
to the spin chain and adding adjoint matter leads to the
elliptic Calogero-Moser (Hitchin type) system. This latter procedure implies
that the first spectral curve (target manifold of the coordinate variables)
is elliptic. At the same time, increasing the dimension of the space-time
leads to the cylindric (5d) or elliptic (6d) second bare curve (target
manifold of the momentum variables). In 5 dimensions the corresponding pure
gauge system is the relativistic Toda chain
\cite{nikita,wdvv,GGM2}\footnote{Note that this system also admits two
representations -- as a spin chain and within the Hitchin-like approach
\cite{kmz}.} and the theory with adjoint matter is the elliptic Ruijsenaars
model \cite{nikita}. Analogous systems in 6 dimensions are described less
manifestly (see, however, \cite{fgnr}), excluding the case of $XYZ$ spin
chain associated with the fundamental matter \cite{GMMM2,GGM2,MM}. At last,
considering theories with the gauge group that is the product of simple
factors, one should enlarge either the matrix dimensions of the Lax operator
at the single site (within the spin chain framework) or the number of marked
points (in the Hitchin-like approach) \cite{GGM1,GGM2}.

I am grateful to A.Gorsky, A.Marshakov and A.Morozov for useful discussions.
I also acknowledge the organizers of the 31st International Simposium
Ahrenshoop on the Theory of Elementary Particles. The work is
partially supported by the grant INTAS-RFBR-95-0690.

\end{document}